\documentclass{article}

\usepackage{arxiv}

\usepackage[utf8]{inputenc} 
\usepackage[T1]{fontenc}    
\usepackage{hyperref}       
\usepackage{url}            
\usepackage{booktabs}       
\usepackage{amsfonts}       
\usepackage{nicefrac}       
\usepackage{microtype}      

\hypersetup{pdfauthor={MJ},pdftitle={Ultrafast MIR Pulse Shaping}}

\usepackage{amsmath}
\usepackage{upgreek}
\usepackage{graphicx}
  \usepackage[dvipsnames]{xcolor} 
\definecolor{myTurquoise}{RGB}{0.0,0.3,0.5}
\definecolor{frenchblue}{rgb}{0.0, 0.45, 0.73}
\definecolor{darkcerulean}{rgb}{0.03, 0.27, 0.49}  
\hypersetup{colorlinks=true,breaklinks=true,linkcolor=darkcerulean,menucolor=darkcerulean,urlcolor=darkcerulean,citecolor=darkcerulean} 

\newcommand{\eg}{e.\,g.}
\newcommand{\myeg}{\eg}
\newcommand{\mycf}{cf.}
\newcommand{\ie}{i.\,e.}
\newcommand{\myie}{\ie}

\newcommand{\figref}[1]{Fig.\,\ref{#1}}
\newcommand{\abkSperrung}[1]{{#1}}
\newcommand{\txt}[1]{\mathrm{#1}}
\newcommand{\CEP}{\abkSperrung{CEP}} 

\newcommand{\DFG}{\abkSperrung{DFG}}

\newcommand{\FTIR}{\abkSperrung{FTIR}}

\newcommand{\FWHM}{\abkSperrung{FWHM}}

\newcommand{\MIR}{\abkSperrung{MIR}}

\newcommand{\OPA}{\abkSperrung{OPA}}
\newcommand{\OPAo}{\OPA\,1}
\newcommand{\OPAt}{\OPA\,2}

\newcommand{\RF}{\abkSperrung{RF}}

\newcommand{\WLG}{\abkSperrung{WLG}}

\newcommand{\AOM}{\abkSperrung{AOM}}

\newcommand{\AWG}{\abkSperrung{AWG}}

\newcommand{\BBO}{\abkSperrung{BBO}}
\newcommand{\KTP}{\abkSperrung{KTP}}

\newcommand{\YAG}{\abkSperrung{YAG}}

\newcommand{\AgGaSe}{A\abkSperrung{gG}\abkSperrung{aS}e$_2$}

\newcommand{\GaSe}{G\textls[20]{aS}e}

\newcommand{\myUspace}{\,}
\newcommand{\myUnitMu}{\upmu}

\newcommand{\mymm}{\myUspace\txt{mm}}
\newcommand{\myumstraight}{\myUnitMu\txt{m}}

\newcommand{\mymu}{\myUspace\myumstraight}
\newcommand{\myum}{\mymu}
\newcommand{\mynm}{\myUspace\txt{nm}}

\newcommand{\myfs}{\myUspace\txt{fs}}

\newcommand{\myps}{\myUspace\txt{ps}}

\newcommand{\mymus}{\myUspace\myUnitMu\txt{s}}
\newcommand{\myus}{\mymus}

\newcommand{\mymuJ}{\myUspace\myUnitMu\txt{J}}
\newcommand{\myuJ}{\mymuJ}
\newcommand{\mymJ}{\myUspace\txt{mJ}}

\newcommand{\mykHz}{\myUspace\txt{\textls[16]{kH}z}}

\newcommand{\myGHz}{\myUspace\txt{\abkSperrung{GH}z}}
\newcommand{\myTHz}{\myUspace\txt{\abkSperrung{TH}z}}

\newcommand{\myrad}{\myUspace\txt{rad}}
\newcommand{\mymrad}{\myUspace\txt{mrad}}

\newcommand{\Tshape}{T_\txt{shape}}

\newcommand{\icomp}{\txt{i}}
\newcommand{\me}{\txt{e}}

\newcommand{\mypi}{\pi} 

\newcommand{\myFT}{\mathcal{FT}}

\newcommand{\mdnuc}{\Delta{}\nu_\mathrm{c}}
\newcommand{\mdw}{\Delta\omega}
\newcommand{\Ecomp}{\tilde{E}}

\title{Generation and Characterization of Tailored MIR Waveforms for Steering Molecular Dynamics}

\author{
	Markus~A.~Jakob\\
	Deutsches Elektronen-Synchrotron, Notkestra{\ss}e 85, 22607 Hamburg, Germany\\
	The Hamburg Centre for Ultrafast Imaging,  Luruper Chaussee 149, 22761 Hamburg, Germany
	\And
	Mahesh~Namboodiri\\
	Deutsches Elektronen-Synchrotron,  Notkestra{\ss}e 85, 22607 Hamburg, Germany
	\And
	Mark~J.~Prandolini\\
	Institut f{\"u}r Experimentalphysik, Universit\"{a}t Hamburg, Luruper Chaussee 149, 22761 Hamburg, Germany\\
	Class 5 Photonics GmbH, Notkestra{\ss}e 85, 22607 Hamburg, Germany
	\And
	Tim~Laarmann\\
	Deutsches Elektronen-Synchrotron,  Notkestra{\ss}e 85, 22607 Hamburg, Germany\\
	The Hamburg Centre for Ultrafast Imaging,  Luruper Chaussee 149, 22761 Hamburg, Germany\\
 	\texttt{tim.laarmann@desy.de} \\
}
\begin{document}
\maketitle
\begin{abstract}
The dream of physico-chemists to control molecular reactions with light beyond electronic excitations pushes the development of laser pulse shaping capabilities in the mid-infrared (\MIR{}) spectral range. Here, we present a compact optical parametric amplifier platform for the generation and shaping of \MIR{}  laser pulses in the wavelength range between $8\myum$ and $15\myum$. Opportunities for judiciously tailoring the electromagnetic waveform are investigated, demonstrating light field control with a spectral resolution of $60\myGHz$ at a total spectral bandwidth of $5\myTHz$. In experiments focusing on spectral amplitude manipulation these parameters result in a time window of $1.8\myps$ available for shaping the temporal pulse envelope and a phase modulation resolution of $100\mymrad$ for several picosecond delays.
\end{abstract}
%
%
\section{Introduction}
Microscopic understanding of chemical dynamics on a molecular level is of fundamental importance in many science disciplines. Current objectives range from unravelling ultrafast transitions in processes relevant to life \cite{Polli2010} to optimizing the efficiency of catalytic processes in material science \cite{Oestroem2015,Buriak2018}. %
%
The gained knowledge will contribute in modern drug design and will help develop materials with novel functionalities \cite{Merz2010,Matzov2017}. Molecular function manifests itself in time-dependent changes of geometric structure, \myie{} bond distances and bond angles describing the molecular dynamics of the system.
With the advent of ultrafast laser pulses in the visible and near-infrared spectral range, the vast majority of spectroscopic studies traced photo-induced processes in electronically excited states \cite{Nibbering2005}. 
However, most chemical reactions of importance to nature and technological applications do not depend on photo excitation. Instead thermal energy and activation barriers govern reaction rates \cite{Stensitzki2018}.
Ultrashort deliberately shaped mid-infrared (\MIR{}) laser pulses 
are well-suited for a systematic investigation of these reactions, because they provide three key properties \cite{Nibbering2005,Gollub2010}. %
%
First, their interaction takes place on the immanent time scales of molecular dynamics. Second, room temperature black body radiation has the maximum emittance around $10\myum$, consequently \MIR{} radiation is well suited to trigger thermal dynamics in a natural environment. Finally deliberate control of the time-frequency distribution of broadband \MIR{} radiation makes it possible to address particular reaction coordinates with a high specificity, \myie{} vibrational modes of interest \cite{Gollub2010}. %
Thereby even reaction pathways can be driven, which are typically not favoured by solely thermally controlled molecular ensembles \cite{LeeSK2012}. In addition, to initiate and monitor molecular dynamics (reactivity/functionality) in space and time requires the formation of quantum mechanical wave packets by coherent coupling of atomic wave functions. Femtosecond \MIR{} laser pulses generated by optical parametric amplification (\OPA) can prepare nuclear wave packets in molecular ensembles avoiding electronic excitation. 
Use of carrier-envelope phase-stable (\CEP-stable) pulses provides control of vibrational wave packets and thereby steering of directed nuclear motion and localization \cite{Thallmair2017,Daoud2018}.
Ultrashort \MIR{} radiation pulses allows to overcome showstopping intramolecular vibrational energy redistribution. For example, if energy can be deposited in specific vibrational modes fast enough, high excitation levels can be reached before the excitation thermalizes by redistributing over all accessible vibrational degrees of freedom \cite{LeeSK2012,Windhorn2003}.
These experimental schemes may be regarded as non chemical (light) stimuli to drive and possibly steer molecular dynamics. In the present contribution we describe and characterize a modular laser infrastructure that has been developed for this purpose.
%
\section{Adjustable 8--15$\boldsymbol{\mymu}$ \OPA{} Design}

\begin{figure} 
	\centering
	\includegraphics[width=0.8\linewidth,trim={5mm 210mm 95mm 0},clip]{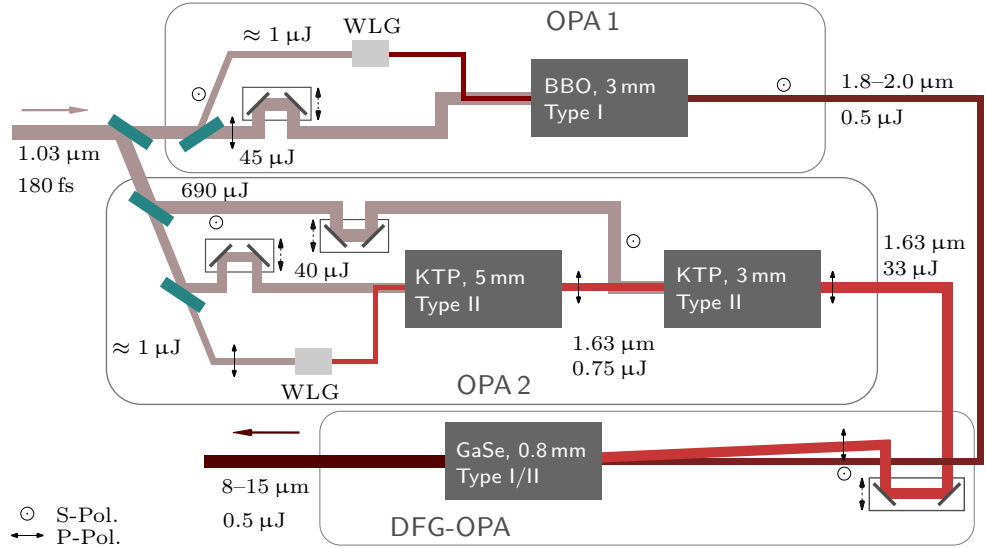}
	\vspace{-6mm}
	\caption{Overview of the compact laser setup. The near-infrared pump laser pulse is converted to the targeted mid-infrared (\MIR) spectral range. White light generation (\WLG), optical parametric amplification (\OPA), and difference-frequency generation (\DFG) are used in several steps.}
	\label{fig:OPAsetup}
\end{figure}
Broadband \CEP{}-stable laser pulses are generated by a compact \OPA{} system similar to the setup described by Sell et al.\ \cite{Sell2008}. The pump laser pulses have an energy of ca.\ $0.8\mymJ$, with a full width at half maximum (\FWHM) duration of ca.\ $180\myfs$, and a wavelength of $1030\mynm$, generated by a commercial Yb-doped potassium gadolinium tungstate (Yb:\abkSperrung{KGW}) optical oscillator and regenerative amplifier. The pump laser can be operated at repetition rates of around $5\mykHz$ with the mentioned pulse energies. Its pulses are converted to \MIR{} pulses whose central wavelength is tunable in the range between 8--$15\myum$. 
The frequency mixing can be performed in typical nonlinear optical crystals, such as \AgGaSe{} and \GaSe{}. These crystals exhibit fairly low linear and two-photon absorption coefficients in the wavelength range between ca.\ $0.7$ and $16\mymu$ \cite{Nikogosyan2005} and have been used extensively in the past \cite{Kaindl2000,Fraser2002,Ventalon2006,Sell2008,Beutler2015}. To avoid significant two-photon absorption by the pump pulses for \MIR{} generation, intermediate \OPA{} conversion steps are necessary. Thereby the final difference frequency generation (\DFG) can be performed with pump wavelengths above $1.4\mymu$ \cite{Kaindl2000,Ventalon2006}. 

\begin{figure} 
	\hspace{16mm}\includegraphics[width=0.8\linewidth]{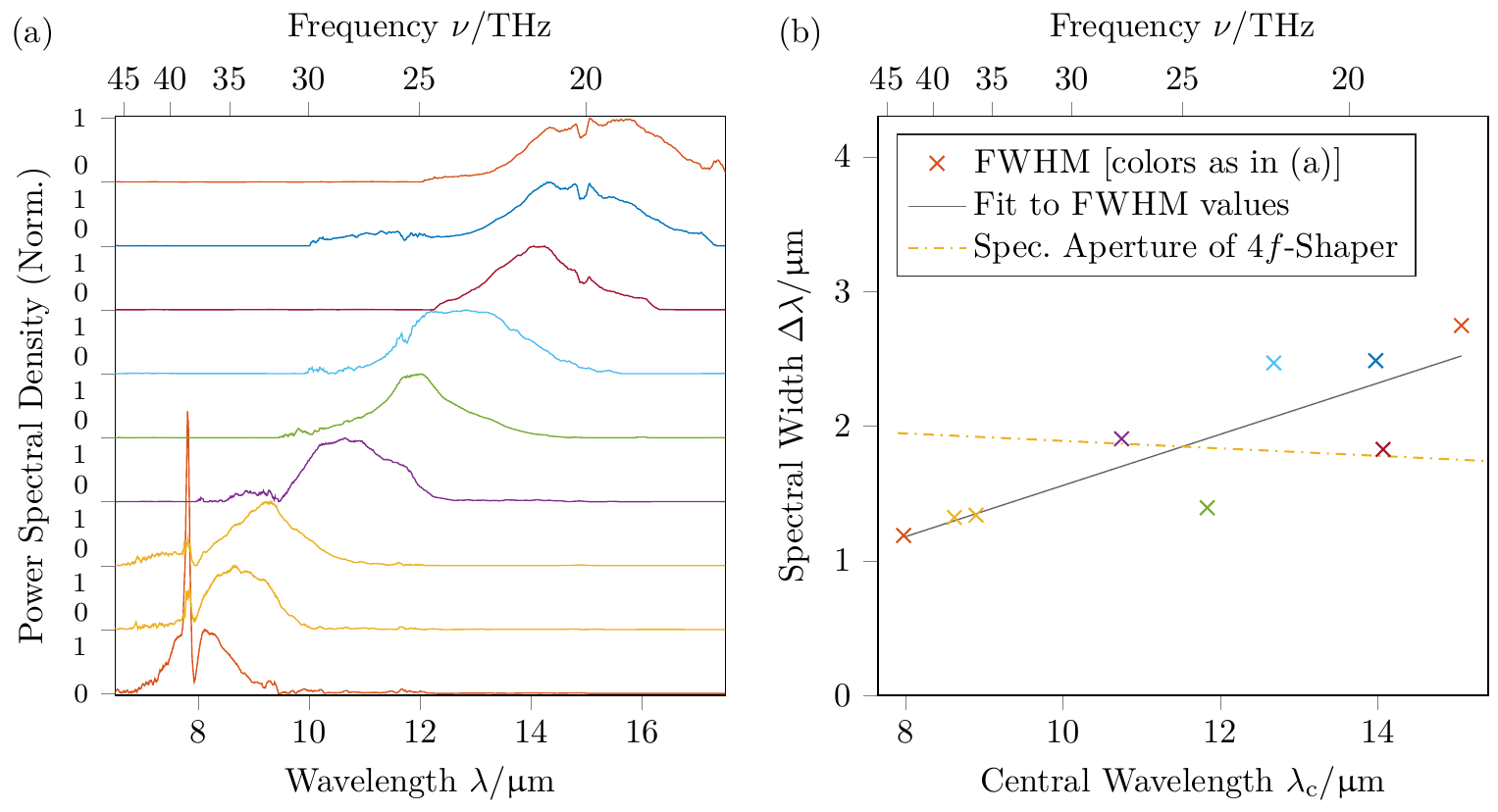}
	\caption{(a) Series of \MIR{} spectra generated with the \OPA{} setup. Artefacts are still visible at around $8\mymu$, emerging from the beamsplitting pellicle used in the \FTIR{} spectrometer and \abkSperrung{CO}$_2$ absorption in air is observed at $15\mymu$. (b) Spectral bandwidth of the respective spectra shown in (a). They are presented including the maximum spectral bandwidth transmitted through the aperture of the acousto-optic modulator mask in the pulse shaper.}
	\label{fig:SpectraAndBandwidth}
\end{figure}

The first intermediate \OPA{} conversion (`\OPAo') 
consists of a single-stage \OPA{} based on $\upbeta$-barium-borate (\BBO{}) and the second of a two-stage \OPA{} (`\OPAt{}') based on potassium titanyl phosphate (\KTP), \mycf{} \figref{fig:OPAsetup}. Two independent white light sources generate the coherent seeds used for each \OPA{}, allowing for independent wavelength tuning \cite{Bradler2009,Bellini2000}. The continua are generated in bulk yttrium aluminium garnet (\YAG) disks of $4$ and $5\mymm$ length.
%
%
\OPAo{}, a single-stage \BBO{} ($3\mymm$, $\theta=22.2^\circ$, type I phase matching) \OPA{}, generates pulses with central wavelengths in between $1.7$--$2.1\myum$, with energies of a few hundred nanojoule up to $0.5\myuJ$ and bandwidths of up to $100\mynm$ at full width at half maximum (\FWHM) of the spectral intensity. Variation of the central wavelength is achieved by changing the angle of the optical axis of the \BBO{} crystal relative to the laser beam path.
%
%
\OPAt{} consists of two \KTP{} stages with $5\mymm$ and $3\mymm$ long crystals, as shown in \figref{fig:OPAsetup}. Type II phase matching is achieved in the \abkSperrung{XZ}-plane of the biaxial crystal under a phase-matching angle $\theta=45.1^\circ$. \OPAt{} generates pulses with pulse energies around $30\myuJ$ and a spectral bandwidth of ca.\ $40\mynm$ at a central wavelength of $1.6\myum$. 
The \MIR{} pulses are finally generated by overlapping the pulses from \OPAo{} and \OPAt{} in a $0.75\myum$ thick \GaSe{} crystal with a small angle of around $1^\circ$. Varying the wavelength of \OPAo{} can be used to choose the frequencies available for frequency mixing in the \MIR{} \OPA{} (`\DFG-\OPA'). Thereby the central wavelength of the \MIR{} pulses can be adjusted in the range between $8$--$15\myum$. Pulse energies reach up to $0.5\myuJ$, at bandwidths of 15--18\%, \myie{} $3.3$--$5.5\myTHz$. Spectra of various generated \MIR{} pulses are presented in \figref{fig:SpectraAndBandwidth}\,(a) along with their spectral bandwidth presented in \figref{fig:SpectraAndBandwidth}\,(b). The pulse duration of the \MIR{}-\OPA{} is estimated to be in the range of $200\myfs$, confirmed by using the \DFG-\OPA{} crystal for a cross-correlation between \OPAo{} and \OPAt{} taking material dispersion into account. The whole \OPA{} architecture is assembled from standard opto-mechanical components and has a compact design fitting on a $90\,\mathrm{cm}$ by $60\,\mathrm{cm}$ breadboard.
%

\section{Ultrafast MIR Pulse Shaping}
\subsection{The AOM MIR Pulse Shaper}
\begin{figure}	 
	\centering
	\includegraphics[width=0.7\linewidth,trim={5mm 220mm 114mm 0},clip]{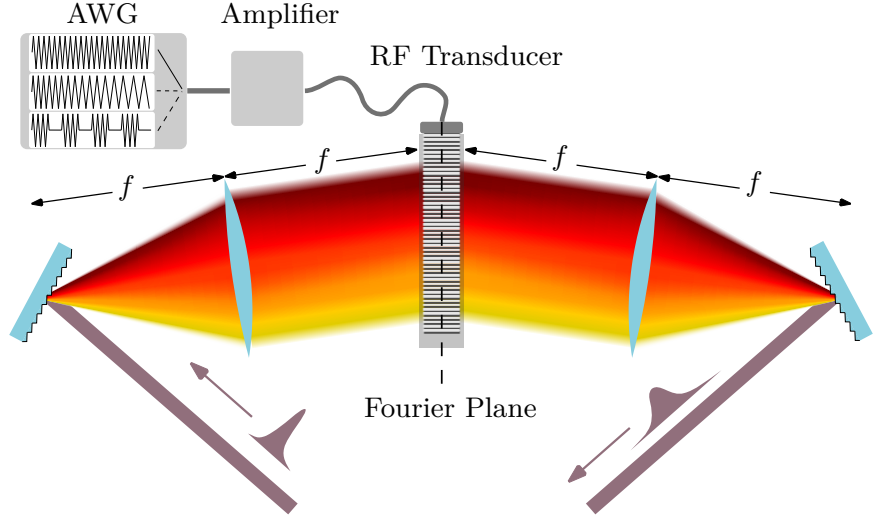}
	\caption{The sketched setup consists of the so called $4f$ configuration, where the acousto-optic modulator mask is placed in the Fourier plane. The arbitrary waveform generator (\AWG), \RF{} amplifier, and \RF{} transducer generate an acoustic transmissive diffraction grating, which modulates the spatially dispersed transmitted broadband radiation.}
	\label{fig:SketchAOMShaper}
\end{figure}

For \MIR{} pulse shaping we used a germanium acousto-optic modulator (\AOM) mask as the active element in a $4f$-configuration, developed in the group of Martin Zanni at University of Wisconsin, Madison \cite{Strasfeld2007}. It allows for the manipulation of the spectral amplitude and phase independently and thereby the time-frequency distribution of individual pulses \cite{Hillegas1994}. This system was adapted to be used in the wavelength range around $10\myum$. The setup is sketched in \figref{fig:SketchAOMShaper}. The active \AOM{} is located in the Fourier plane of the $4f$-configuration. An arbitrary waveform generator (\AWG) generates \RF{} waveforms, which drive a transducer creating a quasi-stationary transmissive diffraction grating. We can therefore modify laser pulses from shot-to-shot, limited in our case by the \RF{} amplifier duty-cycle, or otherwise by the $10\myus$ propagation time of the acoustic mask over the full crystal length. Figure\,\ref{fig:PulsesAndPulseTrainsSimulated} gives examples of general pulse shapes that can be generated.
\subsection{Shaping Window and Pulse Trains}
\begin{figure} 
	\centering
	\includegraphics[width=0.8\linewidth,trim={1mm 0 0 0},clip]{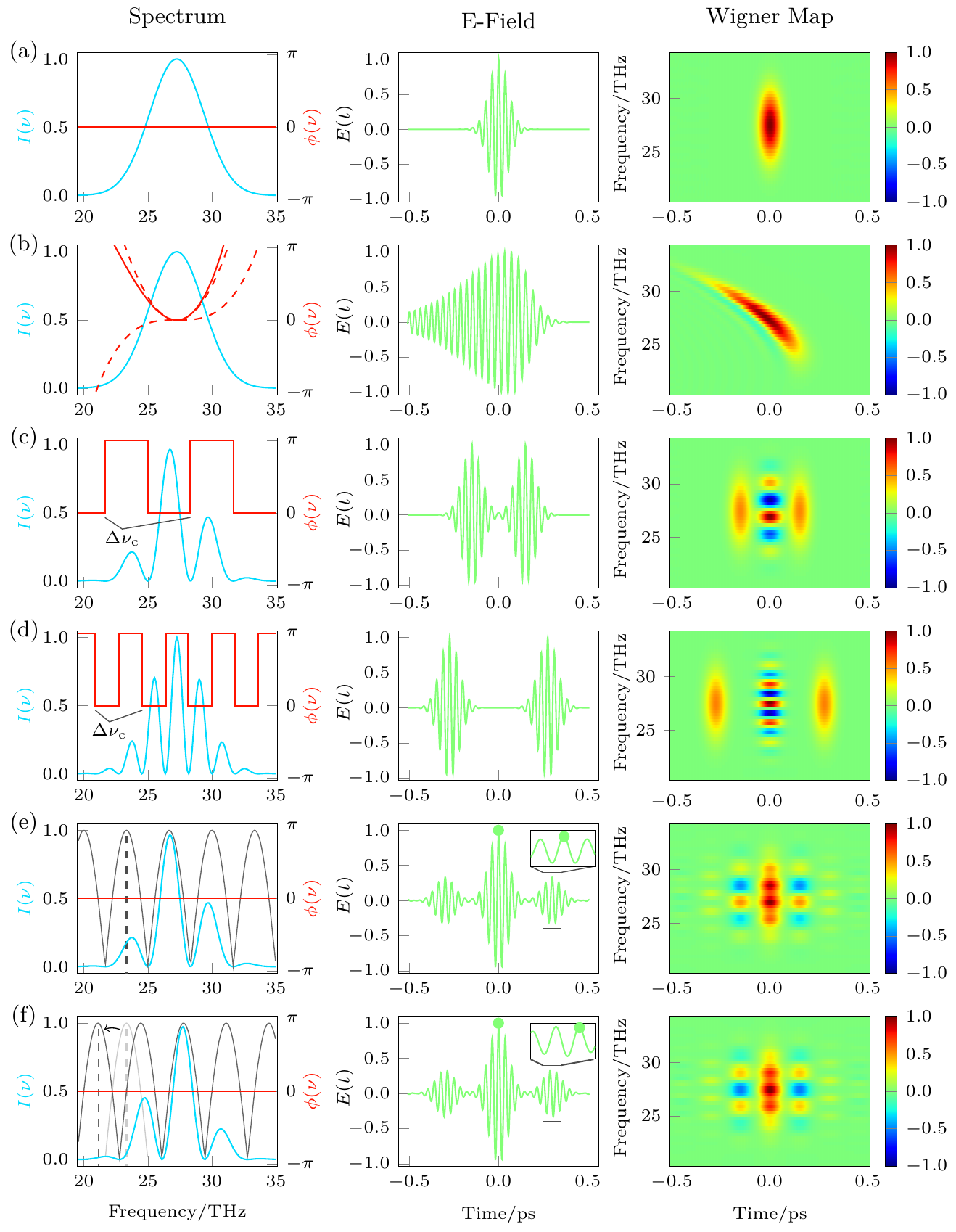}
	\caption{From left to right: spectral intensity $I(\nu)$ including spectral phase $\phi(\nu)$; electric field in the time domain; and Wigner function. Basic pulse shapes are simulated, demonstrating (a,b) the effects of Taylor expansion terms of the spectral phase $\phi(\omega) = \frac{1}{2}\phi^{(2)}\omega^2 + \frac{1}{6}\phi^{(3)}\omega^3$ (dashed, $\phi^{(2)}$ and $\phi^{(3)}$); (c,d) effects of sinusoidal modulation of spectral phase and amplitude with decreasing spectral modulation period $\Delta\nu_\txt{c}$; (e,f) shift of an amplitude shaping mask (gray line) and its effect on \CEP{} of pulses in a pulse train. The pulse train/double pulses demonstrate the reciprocal relationship between spectral modulation and temporal features.}
	\label{fig:PulsesAndPulseTrainsSimulated}
\end{figure}

Real world optical pulse shaping setups impose certain constraints to pulse shaping capabilities. Spectral resolution is in the present case constrained by the diffraction limited beam waist in the Fourier plane. Yet, the accessible spectral resolution is by itself not enough to determine the maximum accessible time window, \myie, the maximum allowed temporal delay between shaped components. The principle of spatial separation of the spectral components of the incoming beam in the $4f$-configuration implies a coupling between temporal delay of frequency components and lateral offset of the outgoing frequency components, \mycf{} Monmayrant et al.\ and references therein. The consequence of this lateral shift may be a deteriorated focus of the beam and hence may affect the application of the beam in experiments \cite{Sussman2008,Frei2009}. In this regard acousto-optic  modulators exhibit an advantage compared to other technical realizations. While liquid crystal (\abkSperrung{LC}) spatial light modulators have fixed pixel gaps in the mask, \AOM{} masks have smooth edges of the effective pixels. Therefore the spatio-temporal coupling only depends on dispersion of the grating and beam size of the incoming beam \cite{Dugan1997,Monmayrant2010}.
%
The following measurement was conducted to examine the size of the pulse shaping window. Pulse trains were used to sample the shaping window of the presented pulse shaping setup. Introducing a comb like structure in the spectral domain leads to pulse trains in the time domain. The separation period $\delta{}t$ depends reciprocally on the frequency modulation period $\Delta\nu$, \myie{} $\delta{}t\propto\Delta\nu^{-1}$. The corresponding effect is simulated in \figref{fig:PulsesAndPulseTrainsSimulated}\,(c) and (d).

\subsubsection{Spectral Resolution and Pulse Separation}
\begin{figure}  
	\centering
	\includegraphics[width=0.8\linewidth,trim={1mm 0 0 0},clip]{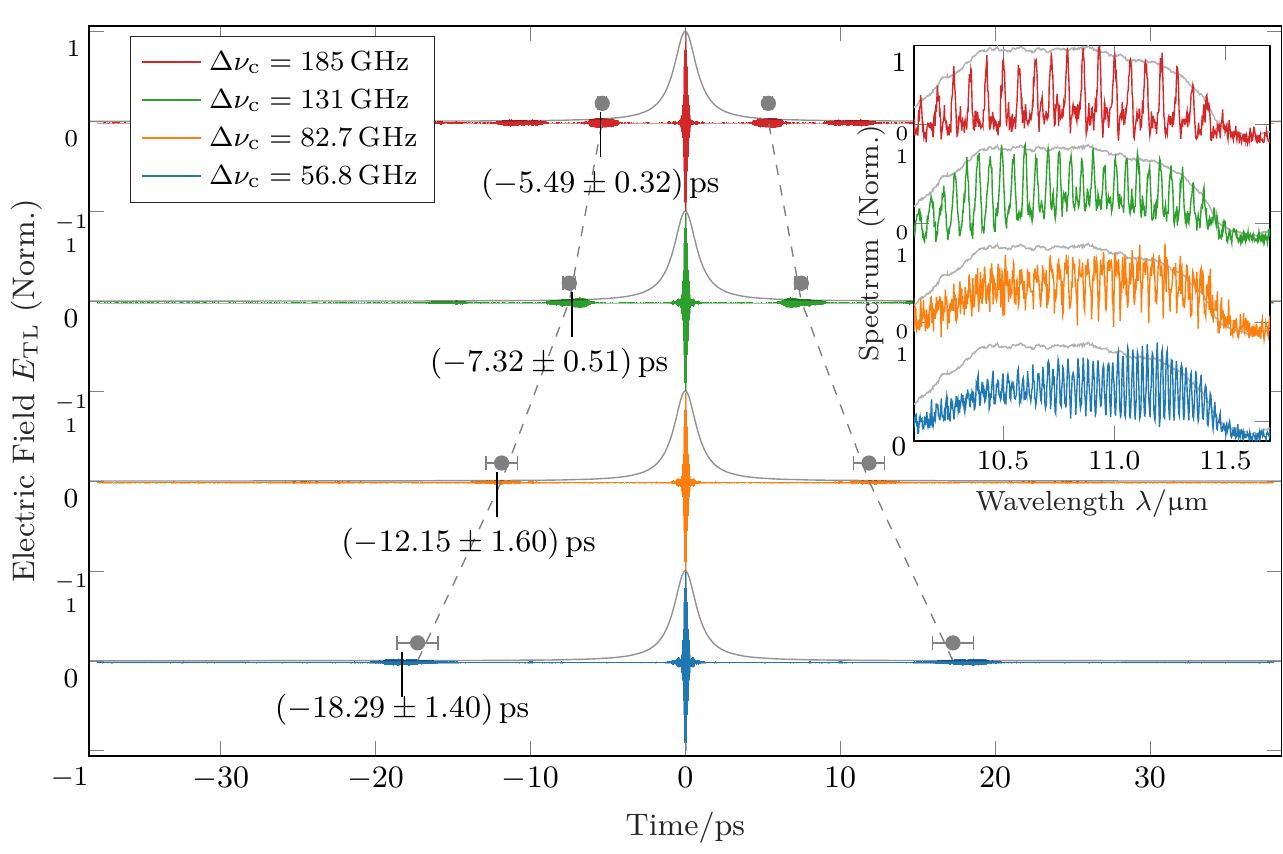}
	\caption{Transform limited electric fields $E_\txt{\abkSperrung{\scriptsize{TL}}}$ reconstructed from comb-shaped spectra (see inset, gray lines represent the non-modified spectrum). The comb spectra generate pulse trains used to sample the wings of the shaping window, a Lorentzian shape shown as gray lines. The expected positions of satellite pulses are indicated by the gray dashed lines. Gray dots and error bars show expected positions according to periodicity in the spectra. Bracketed values show the actual temporal position including standard deviation.}
	\label{fig:MeasuredPulseTrainSeparation}
\end{figure}

The pulse shaper was used to create comb-like amplitude masks, sequentially reducing the spacing $\mdnuc$ of an on-off amplitude pattern. The inset in \figref{fig:MeasuredPulseTrainSeparation} shows the spectra, measured with a Fourier-transform infrared (\FTIR) spectrometer. The spectra shown in \figref{fig:MeasuredPulseTrainSeparation} exhibit peaks of an average width of $23\mynm$, equal to $59\myGHz$ in frequency domain. This value is exceeded for the lowest spectrum in the inset of \figref{fig:MeasuredPulseTrainSeparation} given in blue, in which comb teeth have a width of only $14\mynm$, corresponding to $37\myGHz$. Because the low resolution of $37\myGHz$ is not achieved systematically, the resolution of the pulse shaping setup is taken to be $59\myGHz$ when being used with an initial laser beam width of $2.45\mymm$.

The spectral intensity $I(\nu)$ was used to reconstruct the transform limited electric field time domain using: 
\begin{equation}
E_{\txt{\abkSperrung{\scriptsize{TL}}}} = \myFT^{-1}\left[\sqrt{I(\nu)}\,\me^{\icomp 2\mypi\phi(\nu)}\right]	\,{},
\end{equation} 
where the spectral phase $\phi(\nu)$ is assumed to be zero over the whole spectral range. Details of the mathematical procedure are described for instance in ref.\ \cite{Lee2008}.
In the measurements in presented \figref{fig:MeasuredPulseTrainSeparation}, the distance between the comb teeth of the acoustic mask was decreased step by step. Therefore, the spectral period $\mdnuc$ decreases from top to the lowest shown spectrum in the inset of the figure. The time domain signals $E_\txt{\abkSperrung{\scriptsize{TL}}}$ show an increasing temporal spacing of the pulses in the pulse trains, as demonstrated in \figref{fig:PulsesAndPulseTrainsSimulated}\,(c) and (d). Figure\,\ref{fig:MeasuredPulseTrainSeparation} shows the position of the satellite pulses according to the spacing $\mdnuc$. The satellite pulses in the pulse train sample a time window given as a the gray solid lines. The satellite pulses are chirped and thereby stretched in time. The chirp is caused by the tilted geometry of the $4f$-configuration due to the Bragg reflection.
The shape of the time window is close to a Lorentzian distribution and determines the intensity envelope of the shaping window $\Tshape$ with an \FWHM{} duration of $1.8\myps$, refer to the gray lines in \figref{fig:MeasuredPulseTrainSeparation}. Temporal separation of the pulses is observed up to delays of ca.\ $18\myps$, being measured roughly $0.5\,$m behind the pulse shaper with an \FTIR{} spectrometer.

\subsubsection{\CEP{} Modulation of Subsequent Pulses}
In contrast to a beam splitting delay unit, such as a Michelson interferometer, pulse sequences from a pulse shaper are in general not generated as replicas of the incoming pulse. Identical replicas are generated by modulation of the spectral amplitude and phase with a cosine function, while arbitrary shaping of the individual pulses is achieved by separate interleaving combs \cite{Tsubouchi2009OptComm,Pestov09}.

\begin{figure} 
	\centering
	\includegraphics[width=0.8\linewidth,trim={1mm 0 0 0},clip]{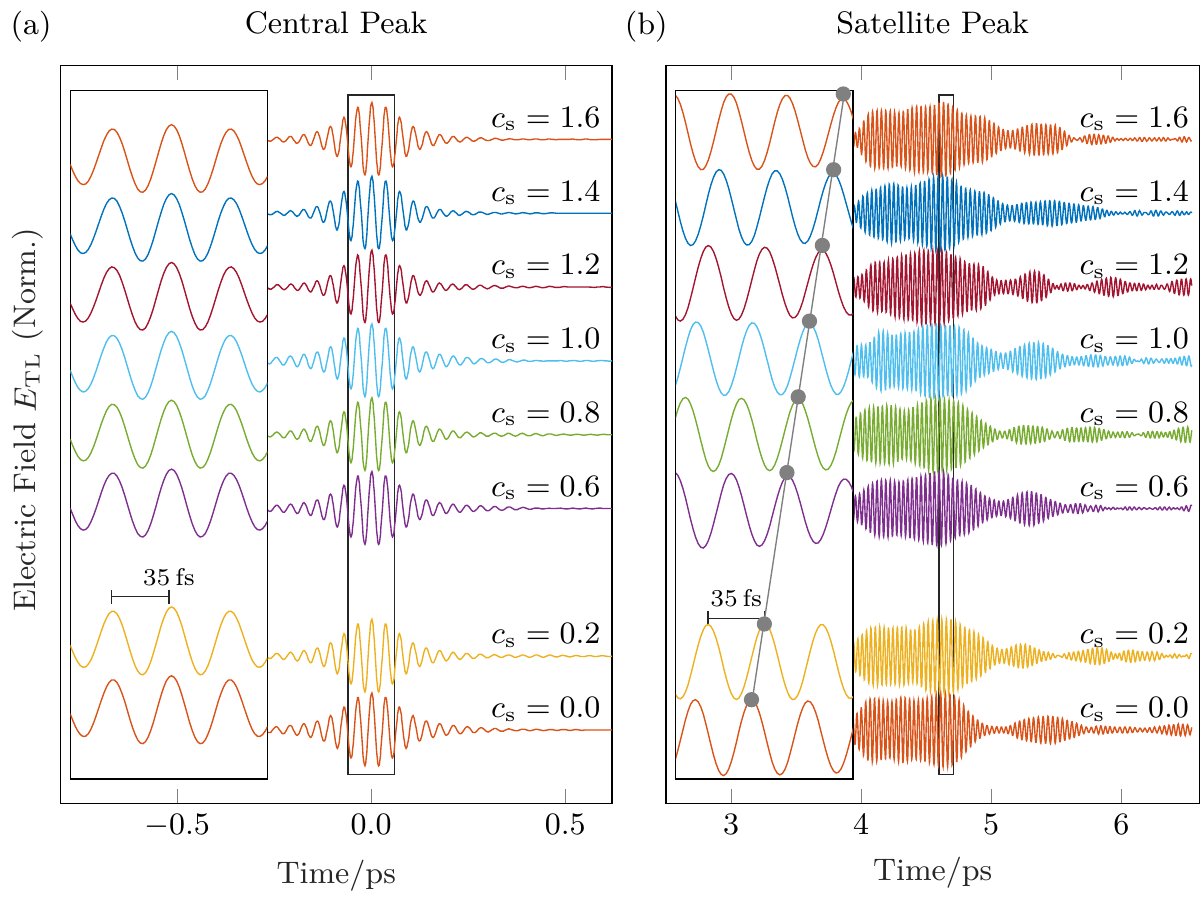}
	\caption{Relative \CEP{} control of a pulse train. Graphs (a) and (b) display electric fields of main and satellite peaks of pulse trains as shown in \figref{fig:PulsesAndPulseTrainsSimulated}\,(e,f). The graph insets highlight the \CEP{} for various positions of a comb shaped mask, which is shifted across the Fourier plane of the $4f$ configuration shaper setup. The according amount of shift of the comb mask is written above the graph lines, in units of one period of the comb pattern $\Delta\varphi=c_\txt{s}2\mypi\myrad$. The gray dots and line in the inset of (b) depicts measurement of the shift in \CEP{} and an according fit. A resolution of better than $100\mymrad$ of \CEP{} relative to the central peak is derived.}
	\label{fig:PulseTrainCEphaseShift}
\end{figure}

In a second measurement amplitude pulse shaping was used to demonstrate \CEP{} modulation, which can be investigated by a simple spectrometer. Transform limited pulse trains are investigated as a test case for \CEP{} modulation of pulse trains. The spectral intensity distribution does not contain information of the phase of the electric field. It can only be assumed that the spectral phase is of a certain shape. A constant spectral phase $\phi(\omega)$ is assumed over the whole spectrum. It is then possible to perform amplitude pulse shaping, thereby modifying the relative \CEP{} of the satellite pulses with respect to the central pulse in the pulse train. According to the Fourier shift theorem a shift in the spectrum by an amount $\mdw_0$ leads to a phase shift in the electric field: 
\begin{equation}
\Ecomp(\omega) \equiv \myFT[\Ecomp(t)] \quad \Rightarrow \quad \Ecomp(\omega-\mdw_0) = \myFT[ \me^{\icomp \mdw_0 t} \Ecomp(t) ] \,.
\end{equation}
The simulation of this approach is shown in \figref{fig:PulsesAndPulseTrainsSimulated}\,(e) and (f) and was pursued in the following experiment.
%
%
%
%
%
%
A comb mask with a fixed periodicity was shifted laterally in the Fourier plane, while the overall envelope of the spectrum stayed constant. The generated pulse trains thereby exhibit a fixed temporal separation. Only the relative \CEP{} of the single pulses are modified. Figure\,\ref{fig:PulseTrainCEphaseShift} shows the central pulses and a satellite pulse of a pulse train as in the previous measurement. The \CEP{} of the central pulse relative to the \CEP{} of the satellite pulses is subsequently modified. The spectral comb pattern was shifted for 1.6 comb periods $\mdnuc$. As \figref{fig:PulseTrainCEphaseShift}\,(a) and (b) depict for transform-limited electric fields $E_\txt{\abkSperrung{\scriptsize{TL}}}$ the \CEP{} of the central pulse was not modified--its \CEP{} is assumed to be $\phi(\omega)\equiv0\myrad$--while the \CEP{} of the first satellite peak shifts proportional to the lateral shift of the comb mask. The stability of the demonstrated relative \CEP{} control in pulse trains is on the order of $100\mymrad$ for pulse separations of $\delta{}t = 4.2\myps$. This value is derived from a fit to the phases and presented in the inset of \figref{fig:PulseTrainCEphaseShift}\,(b).

\section{Conclusion}
In the present paper we demonstrate a compact \OPA{} setup generating sub-microjoule pulses in the long wavelength \MIR{} covering 8--$15\mymu$. The spectral bandwidth supports few-cycle pulses, while the \OPA{} architecture delivers passively \CEP{}-stable pulses. The spectral bandwidth of the pulses perfectly fits the spectral bandwidth that can be accessed by the implemented $4f$-configuration \AOM{} pulse shaper. The pulse shaping capabilities of the setup are demonstrated for amplitude pulse shaping around $10.5\myum$. The reconstruction of the transform limited electric fields is used to determine the available time window for pulse shaping of $1.8\myps$. It is planned to extend the window in the near future by increasing the beam size of the \MIR{} beam sent into the pulse shaper setup from $2.45\mymm$ to more than $7\mymm$ and thereby alleviate effects of lateral beam offsets due to spatio-temporal coupling discussed in the present contribution.
Control of the relative \CEP{} of subsequent pulses in a pulse train is found to be better than $100\mymrad$ for delays in the range of several picoseconds. Table\,\ref{tab:ShaperProperties} summarizes the above evaluated values along with theoretical estimations. These are relevant for applications using the shaper platform with a total transmission of 10\% to steer chemical dynamics in the electronic ground state.

\begin{table}\centering
	\caption{Summary of the pulse shaper properties for the application at $10.8\mymu$. It provides intensity \FWHM{} of total transmitted spectral width $\Delta\lambda$, the spectral resolution $\delta\nu$, the shaping window $\Tshape=4\ln(2)/\delta\omega$, the number of effective `pixels' $n$, derived from optical resolution over full bandwidth, and the achieved complexity $\eta=\Delta\lambda/\delta\lambda$. The calculated parameters are derived from the beam radius, according to equations as demonstrated \myeg{} in ref.\,\cite{Monmayrant2010}. The measured values are derived from the presented measurements. Values given in brackets indicate the best values that may potentially be achieved.
	Starred ($^*$) values are calculated assuming a Gaussian spectrum.
	}%
			\begin{tabular}{ l c c}
				\toprule
				Parameter        & Calculation              & Measurement            
				\\ \midrule
				$\Delta\lambda$  & $1.9\mymu$ 
				& $1.9\mymu$             
				\\
				$\delta\lambda$      & $20\mynm$              & $23\mynm$ ($14.2\mynm$) 
				\\
				$\delta\nu$      & $55\myGHz$              & $59\myGHz$ ($37\myGHz$) 
				\\
				$T_\txt{shape}$  & $1.8\myps$              & $7.5\myps^*$          
				\\
				$n_\txt{pixels}$ & $152$                   & 142$^*$ (262$^*$)      
				\\
				$\eta$           & 90                      & 84$^*$ (133$^*$)       
				\\  		\bottomrule	
		\end{tabular}   
	\label{tab:ShaperProperties}%
\end{table}%
%
%
%
%
\section*{Funding}
Cluster of Excellence `The Hamburg Centre for Ultrafast Imaging' of the Deutsche
Forschungsgemeinschaft (\abkSperrung{DFG}) - \abkSperrung{EXC} 1074 - project \abkSperrung{ID} 194651731.
\section*{Acknowledgments}
We thank Giulio M.\ Rossi, Robert Riedel, and Michael Schulz for fruitful discussions.
\section*{Disclosures}
The authors declare that there are no conflicts of interest related to this article.
\bibliographystyle{osajnl}  
\bibliography{bib_file_main,bib_file_books}

\end{document}